\documentclass[prl,twocolumn,floatix,amssymb,amsmath,superscriptaddress]{revtex4-1}
\usepackage{graphicx}

\begin{document}

\newcommand{\ket}[1]{\ensuremath{\left|{#1}\right\rangle}}
\newcommand{\bra}[1]{\ensuremath{\left\langle{#1}\right|}}
\newcommand{\quadr}[1]{\ensuremath{{\not}{#1}}}
\newcommand{\quadrd}[0]{\ensuremath{{\not}{\partial}}}

\title{Klein tunneling and Dirac potentials in trapped ions}
\author{J. Casanova}
\affiliation{Departamento de Qu\'{\i}mica F\'{\i}sica, Universidad del Pa\'{\i}s Vasco - Euskal Herriko Unibertsitatea, Apdo.\ 644, 48080 Bilbao, Spain}
\author{J.~J. Garc{\'i}a-Ripoll}
\affiliation{Instituto de F\'{\i}sica Fundamental, CSIC, Serrano 113-bis, 28006 Madrid, Spain}
\author{R. Gerritsma}
\affiliation{Institut f\"ur Quantenoptik und Quanteninformation, \"Osterreichische Akademie der Wissenschaften, Otto-Hittmair-Platz 1, A-6020 Innsbruck, Austria}
\affiliation{Institut f\"ur Experimentalphysik, Universit\"at Innsbruck, Technikerstrasse 25, A-6020 Innsbruck, Austria}
\author{C. F. Roos}
\affiliation{Institut f\"ur Quantenoptik und Quanteninformation, \"Osterreichische Akademie der Wissenschaften, Otto-Hittmair-Platz 1, A-6020 Innsbruck, Austria}
\affiliation{Institut f\"ur Experimentalphysik, Universit\"at Innsbruck, Technikerstrasse 25, A-6020 Innsbruck, Austria}
\author{E. Solano}
\affiliation{Departamento de Qu\'{\i}mica F\'{\i}sica, Universidad del Pa\'{\i}s Vasco - Euskal Herriko Unibertsitatea, Apdo.\ 644, 48080 Bilbao, Spain}
\affiliation{IKERBASQUE, Basque Foundation for Science, Alameda Urquijo 36, 48011 Bilbao, Spain}

\begin{abstract}
We propose the quantum simulation of the Dirac equation with potentials, allowing the study of relativistic scattering and the Klein tunneling. This quantum relativistic effect permits a positive-energy Dirac particle to propagate through a repulsive potential via the population transfer to negative-energy components. We show how to engineer scalar, pseudoscalar, and other potentials in the $1+1$ Dirac equation by manipulating two trapped ions. The Dirac spinor is represented by the internal states of one ion, while its position and momentum are described by those of a collective motional mode. The second ion is used to build the desired potentials with high spatial resolution.
\end{abstract}

\maketitle

The Dirac equation describes the successful merge of quantum mechanics with special relativity~\cite{Thaller92}, predicting the electron spin and antimatter. Surprisingly, it also predicts some conflictive quantum relativistic effects, as the {\it Zitterbewegung}~\cite{Schrodinger30} and the Klein paradox~\cite{Klein29}, which have been discussed theoretically though never experimentally tested.  Presently, there is growing interest in different aspects of quantum simulations~\cite{Aspuru-Guzik05,Friedenauer08}. Recently, the quantum simulation of a free-particle Dirac equation in trapped ions~\cite{Lamata07,Gerritsma10a} has generated a dialogue between relativistic quantum mechanics and quantum optics~\cite{Bermudez07a,Bermudez07b}. Here, we propose the quantum simulation of the Dirac equation with potentials in trapped ions, allowing us to explore the unintuitive physics of relativistic scattering, especially when compared with Schr\"odinger quantum mechanics. We also show that in $1+1$ dimensions only the scalar and pseudoscalar potentials can be used to confine a Dirac particle. Moreover, we discuss how electromagnetic potentials give rise to Klein tunneling~\cite{GrapheneKP}, in which a particle propagates through a repulsive potential by turning into its antiparticle~\cite{Krekora04}.

The Dirac equation with a covariant potential~\cite{Thaller92} can be written as
\begin{equation}
  \label{eq:dirac-with-V}
  \left[-i\hbar\quadrd - mc - \frac{1}{c}V_{cov} \right] \psi = 0 ,
\end{equation}
where we used Feynman's notation, $\quadr{A} = \gamma^\mu A_\mu$, being $\gamma^\mu$ the Dirac matrices and $A_\mu$ a four-vector. The external potential $V_{cov}$ can take many forms that transform differently under Lorentz rotations and boosts. In particular, $V_{cov}$ includes scalar potentials that add to the relativistic mass term, as well as pseudoscalar, electric and magnetic potentials, among other cases. Multiplying Eq.~(\ref{eq:dirac-with-V}) by $\gamma^0 :=\beta,$ and introducing the vector of matrices $\vec{\alpha} = \gamma^0\vec{\gamma},$ the momentum operator $\vec{p}=-i\hbar\nabla,$ and the time $t=x_0/c,$ we obtain
\begin{equation}
  i\hbar \partial_t \psi =
  \left[c\vec{\alpha}.\vec{p} + mc^2\beta + \beta V_{cov} \right] \psi .
\end{equation}
Up to unitary transformations, there are different sets of matrices, $\alpha$ and $\beta$, determining different Dirac representations.
In the case of $1+1$ dimensions, in particular, $\alpha$ and $\beta$ can be chosen as any two different Pauli matrices. For simplicity, we will use a real representation,
\begin{equation}
  \label{eq:dirac1+1}
  \gamma^0=\beta=\sigma_z,\;\gamma^1=i\sigma_y,\;\alpha=\sigma_x.
\end{equation}

There are six ways to introduce a potential in the Dirac equation~\cite{Thaller92}, depending on its behavior under Lorentz transformations. Using the pseudoscalar operator $\gamma^5:=i\prod_\mu \gamma^\mu$ and the tensor $\sigma^{\mu\nu}:=i\gamma^\mu\gamma^\nu,$ these are
\begin{eqnarray}
  \label{eq:dirac-pot}
  V_{cov} &=& V + q\quadr{A} + B_{\mu\nu}\sigma^{\mu\nu}+ q \gamma^5 \tilde{V} + \\
  &+& q \gamma^5 \tilde{\quadr{A}} +
  \tilde{B}_{\mu\nu}\gamma^5 \sigma^{\mu\nu}.\nonumber
\end{eqnarray}
The potential $V$ transforms as a scalar and mimics an induced mass term. The field $A$ transforms as a four-vector and corresponds to the electromagnetic potential acting on a charge $q,$ with the electric potential $A_0=\phi$ and a three-vector component $\vec{A}$. The remaining four potentials $B, \tilde{A}, \tilde{V}$ and $\tilde B$, transform as matrices, pseudoscalars, pseudovectors, and pseudotensors, respectively, and behave as anomalous field moments~\cite{Thaller92}.

In $1+1$ dimensions the landscape simplifies considerably and we can choose a parametrization such that
\begin{eqnarray}
  i\hbar\partial_t\psi &=& \left[
    c \sigma_x\left(p - \frac{q}{c}A\right) + q \phi
    + (mc^2 + V)\sigma_z - q\tilde V\sigma_y \right] \psi \nonumber\\
  &=& H(q) \psi , \label{H-eff}
\end{eqnarray}
where only  $A$, $\phi$, $V$, and $\tilde V$ are nonzero. We will first focus on potentials that are linear in the particle position
\begin{equation}
  V = \upsilon_{sc}x,\; \phi = \upsilon_{el}x/e,\; A= \upsilon_{mag}x/e,\;\mathrm{and}\;
  \tilde{V} = \upsilon_{ps}x/e,\label{potentials}
\end{equation}
where $e$ is the unit charge, and we will assume $q=\pm e.$

All the preceding potentials can be simulated with trapped ions~\cite{Comment}. Let us consider a string of two trapped ions, 1 and 2, which could be $^{40}$Ca$^+$ ions with long-lived internal states $|S_{1/2},m=1/2\rangle$ and $|D_{5/2},m=3/2\rangle$. The first ion will encode a Dirac spinor in those internal states, while the second ion will be used as an ancilla to implement potentials. Assuming the validity of the Lamb-Dicke approximation, addressed laser beams can be used to implement interactions of the forms $H_c=\hbar\tilde{\Omega}(\sigma^+_je^{i\phi}+\sigma^-_je^{-i\phi})$ (carrier), $H_r=\hbar\eta\tilde{\Omega}_r(a\sigma^+_je^{i\phi_r}+a^{\dag}\sigma^-_je^{-i\phi_r})$ (red sideband) and $H_b=\hbar\eta\tilde{\Omega}_b(a^{\dag}\sigma^+_je^{i\phi_b}+a\sigma^-_je^{-i\phi_b})$ (blue sideband) on each ion $j=1,2$. Here, $\tilde{\Omega}_{(b,r)}$ and $\phi_{(b,r)}$ are the Rabi frequency and phase of each light field, $\eta \ll 1$ is the Lamb-Dicke parameter and $\sigma^+_j$ ($\sigma^-_j$) and $a^{\dag}$ ($a$) are raising (lowering) operators for the ion's internal states and for a collective motional mode. For the one-dimensional case, we could use, for instance, the center-of-mass mode.

As shown in~\cite{Lamata07,Gerritsma10a}, the Dirac Hamiltonian for a free particle, $H^{\rm free}_D = c\sigma_x\hat{p} + mc^2\sigma_z$, can be engineered using a simultaneous blue and red sideband interaction of equal strength with the appropriate relative phase, amplitude and detuning $\hbar\Omega := mc^2$. Here, $\hat{p}=i\hbar\frac{a^{\dag}-a}{2\Delta},$ $\Delta=\sqrt{{\hbar}/{4\tilde{m}\omega}}$ is the size of the ground state wave packet, $\tilde{m}$ is the mass of a single ion and $\omega$ is the trap frequency. We consider the laser coupled only to the internal state of ion 1 (see Fig.~1) and show how to extend this model to include each of the potentials $V$, $\tilde{V}$, $A$, and $\phi$.

\begin{figure}[t]
  \centering
  \includegraphics[width=0.35\linewidth]{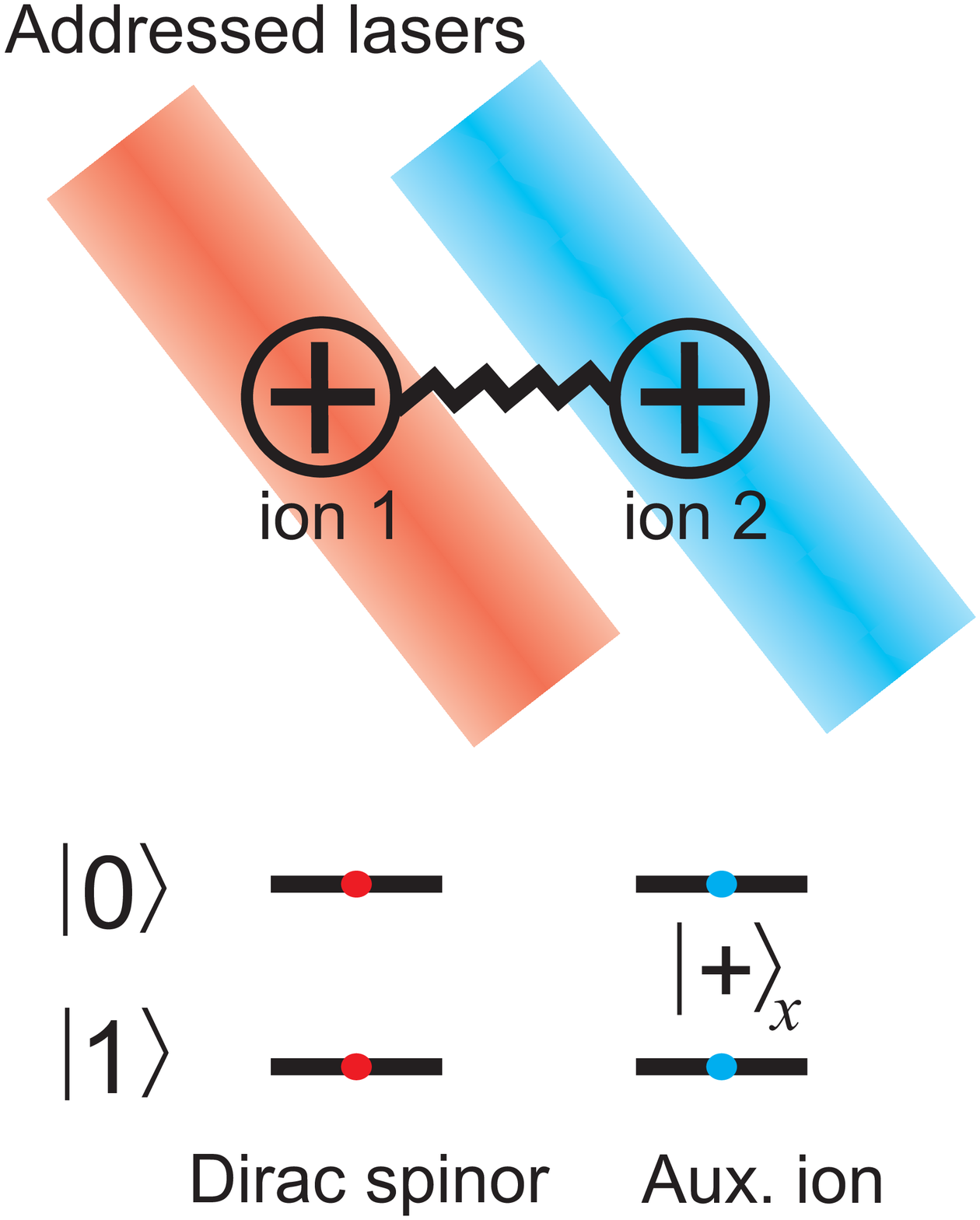}
  \caption{Setup for simulating a Dirac particle in an external potential. A string of two trapped ions is manipulated with addressed laser beams, coupling to each ion's internal state and the collective motion. The Hamiltonian for a free Dirac particle and the potentials $A$, $V$ and $\tilde{V}$ can be implemented directly by the laser light impinging on ion 1. The electrostatic potential $\phi$ is implemented by red and blue sidebands applied to the auxiliary ion to create an interaction $\propto\sigma^x_{2}\hat{x}$, with the auxiliary ion prepared in an eigenstate of $\sigma^x_2$.}
  \label{fig:klein}
\end{figure}

The potentials $A$ and $\tilde V$ can be created simultaneously by changing the phase and the intensity of the sidebands. Note that the total Hamiltonian $(c\sigma_x \hat{p} - e \sigma_x \upsilon_{mag} \hat{x} - e \upsilon_{ps} \hat{x} \sigma_y)$, with $\hat{x}=(a^{\dag}+a)\Delta$, can be decomposed in terms of $a^\dag \sigma^{\pm},$ and $a\sigma^{\pm}$. For the implementation of the scalar potential, we can use two additional light fields acting on ion 1 with Rabi frequency $\tilde{\Omega}_{sc}$, one blue detuned by $\omega/2$ from the spinor transition, the other red detuned by the same amount to create a Hamiltonian term $4\hbar\eta\Omega_{sc}^2\hat{x}\sigma^z_1/\omega\Delta$~\cite{Kim08}. Alternatively, we can choose another representation in which the mass term acquires the Pauli matrix $\sigma_y$. In this case, the term $mc^2$ is created by a carrier excitation, whereas the scalar potential can be engineered from blue and red sidebands in the same way as $A$ and $\tilde V$. Finally, the electrostatic field can be implemented by driving a blue and red sideband simultaneously on the auxiliary ion. With appropriate phases the resulting Hamiltonian term becomes $\propto\sigma^x_{2}\hat{x}$. Preparing the auxiliary ion in an eigenstate of $\sigma^x_{2}$, this operator can be replaced by its eigenvalue and the interaction reduces to the desired form. A general Hamiltonian for this system that can be compared to Eq.~(\ref{H-eff}), in the rotating frame and after rotating-wave approximation, reads
\begin{eqnarray}
\label{IonModel}
H = && \, \hbar \eta \left(\tilde{\Omega}_b e^{i\phi_b} a^\dag \sigma^+_1 + \tilde{\Omega}_r e^{i\phi_r} a \sigma^+_1 +  \mathrm{H.c.} \right) \nonumber \\
&& + \hbar\eta\tilde{\Omega}_2 \sigma^x_{2} \hat{x}/\Delta + \hbar(\Omega + 4 \eta\Omega_{sc}^2 \hat{x}/\omega\Delta ) \sigma^z_1 . 
\end{eqnarray}
Here, the first two terms with their Hermitian conjugates describe blue and red motional sidebands, the term involving the second ion describe a conditional displacement, and the last term involves constant and position-dependent Stark shifts. The relations between the Dirac model in Eq.~(\ref{H-eff}) and the ion system in Eq.~(\ref{IonModel}) are
\begin{eqnarray}
  \hbar\Omega = mc^2,  \,\,\, \hbar\eta\tilde{\Omega}_2/\Delta  = \upsilon_{el}, \,\,\,
  4\hbar\eta\Omega_{sc}^2/\omega\Delta = \upsilon_{sc} ,
\end{eqnarray}
while the relative weights of $c$, $\upsilon_{mag}$, and $\upsilon_{ps}$, can be set by changing phases $\phi_{b,r}$ and amplitudes $\Omega_{r,b}$.

The creation of an electrostatic potential via a detuned laser acting on ion $2$ yields the interesting case of interaction terms $\hbar \eta\tilde{\Omega}_2\hat{x}\sigma^x_2/\Delta+\hbar \Omega_2\sigma^z_2$. For a large detuning $\Omega_2 \gg \eta \tilde{\Omega}_2$ this interaction becomes effectively $\propto \hat{x}^2 \sigma^z_2$, allowing the simulation of quadratic potentials~\cite{giachetti09}.

\paragraph{Electric potential and Klein tunneling.-} Not all potentials can actually confine a Dirac particle. To prove this let us introduce an antiunitary operation known as charge conjugation $\psi_c = K \psi = C \psi^\star,$ combining complex conjugation and a unitary matrix $C\gamma^{\mu\star}C^{-1}=-\gamma^\mu.$ The charge conjugate spinor satisfies a Dirac equation with opposite charge $i\hbar\partial_t\psi_c = H(-q)\psi_c,$ in which certain terms of the covariant potential which are proportional to the charge ``q'' have changed sign. We can also show that the negative energy branch of a Dirac equation can be antiunitarily related to the positive energy branch with opposite charge $K H(q) K^{-1} = -H(-q).$ 
In other words, the negative energy branch of the Dirac spectrum are indistinguishable from particles with opposite charge. Moreover, these antiparticles will see certain components of the covariant potential, $A, \tilde V, \tilde A,$ with opposite sign as for their positive energy counterparts.

The fact that positive and negative energy states see different effective potentials, and that both components are coupled, allows the apparition of the so-called Klein paradox~\cite{Thaller92}. The electric potential in Eq.~(\ref{potentials}) is repulsive for positive-energy charged particles in a large region, $(q/e)\upsilon_{el} x > 0,$ but it will allow the antiparticle or negative energy states tunnel into a region that would be forbidden in the (Schr\"odinger)  nonrelativistic regime (see Fig.~\ref{fig:klein}). The situation is even more interesting, for if we send a massive charged positive-energy particle against an electrostatic potential barrier and the particle has enough energy, it will split into a positive and negative wavepackets, the former bouncing back and the latter penetrating into the energy barrier (see Fig.~\ref{fig:scattering}).

\begin{figure}[t]
  \centering
  \includegraphics[width=1\linewidth]{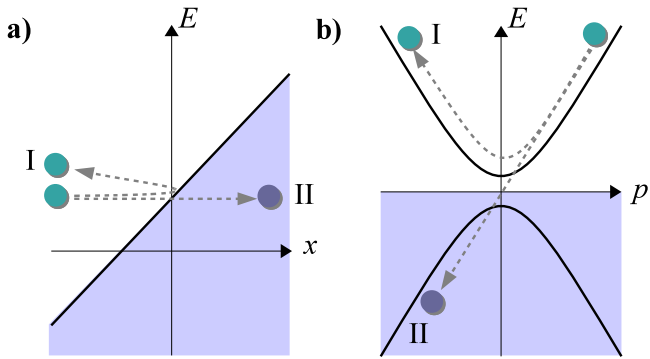}
  \caption{Klein tunneling in a repulsive potential, $\phi_{el}(x)= \upsilon_{el}x.$ (a) In position space, the particle may bounce back (I) or enter the forbidden region by reducing its kinetic energy (II). (b) In momentum space, (I) and (II) correspond to the particle having positive energy or turning into an antiparticle.}
  \label{fig:klein}
\end{figure}

We explain now how Klein tunneling can be interpreted in terms of a quantum optical concept: the Landau-Zener tunneling~\cite{Sauter31}. Working in momentum space, where $x=(+i\hbar\partial_p),$ and with charge $q=e,$ the electric potential $\upsilon_{el}x$ is equivalent to a deceleration or a decrease in the particle momentum, which can be compensated using the change of variables $\psi(p,t) = \xi(p + \upsilon_{el}t)$,
\begin{equation}
  \label{eq:landau-zener}
  i\hbar \partial_t\xi = [c \sigma_x (p - \upsilon_{el}t) + mc^2 \sigma_z]\xi.
\end{equation}
This equation corresponds exactly to a Landau-Zener process in which an effective magnetic field along the $x$ direction is increased linearly in time.

\begin{figure*}[t]
  \includegraphics[width=0.95\linewidth]{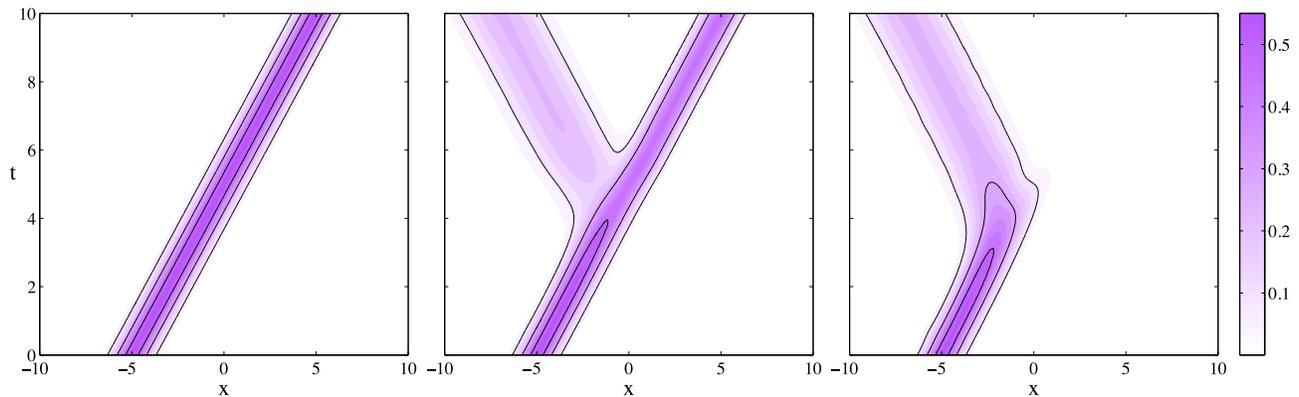}
  \caption{Scattering events for three effective masses, $m=0,0.5$ and $1$ (left to right). We plot the total probability density as a function of space $x$ and time $t.$ We have used natural units $c=\hbar=1,$ and a fixed potential slope $c \upsilon_{el}=1.$}
  \label{fig:scattering}
\end{figure*}

The dynamics is summarized in Fig.~\ref{fig:klein}. If the particle initially moves against the slope, that is $p>0,$ the potential will decelerate the particle which may (I) bounce back or (II) tunel inside the potential barrier. The first case corresponds to adiabatic transition in the Landau-Zener picture, it happens when the change of the momentum is small, $\hbar{c}|\upsilon_{el}| \ll m^2c^4,$ and is the typical behavior of nonrelativistic particles. However, for large enough slope or in the relativistic limit, it becomes possible for the particle to switch branch, acquiring negative kinetic energy and entering the originally forbidden region. More precisely, the probability of transition can be computed with the Landau-Zener formula as follows
\begin{equation}
  P_{II} = \exp\left(-2\pi\frac{m^2c^4}{2 \hbar c\upsilon_{el}}\right).
\end{equation}
Note that once the particle has switched to negative energy states, it has an opposite charge and the potential $\phi_{el}$ accelerates the particle indefinitely in the opposite direction. This makes such a potential effectively a filter that separates particles from antiparticles.

Using Eq.~(\ref{eq:landau-zener}), it is easy to perform numerically accurate simulations of the Klein scattering for arbitrary initial conditions and different ratios of the potential slope, $m^2c^4/\hbar c\upsilon_{el}.$ In Fig.~\ref{fig:scattering}, we see that for very small masses the particle is either not confined, or it splits into a negative charge and a positive charge components [Fig.~\ref{fig:scattering}b]. In the nonrelativistic limit, $m^2c^4 \gg \hbar c \upsilon_{el},$ though, the particle mostly bounces back from the barrier. The first plot corresponds only to situation II, the second plot combines I and II, and finally the third situation is just I.

\paragraph{Scalar potential.-} As we discussed above, the Klein paradox only applies to potentials which are not invariant under charge conjugation. A scalar potential~\cite{Hiller02} such as $V$ in Eq.~(\ref{eq:dirac-pot}), acting similarly on both particle and antiparticle states, can still confine and reproduce the physics we are used to in the nonrelativistic limit. In particular for our choice of linearly growing potential $V = \upsilon_{sc}x,$ with $\upsilon_{el},\upsilon_{mag},\upsilon_{ps}=0,$ it is possible to show the existence of bound orbits. This is best analyzed by squaring the effective Hamiltonian,
\begin{equation}
  H^2 = cp^2 + (\upsilon_{sc}x + mc^2)^2 + c\hbar \upsilon_{sc}\sigma_y.
\end{equation}
The first implication is that the scalar potential eigenenergies correspond to those of a harmonic oscillator
\begin{equation}
  E^2 = 2\hbar c \upsilon_{sc} \left (n + \frac{1}{2} \pm 1\right).
\end{equation}
The second implication is that the conserved quantity $H^2$ is defining elliptical orbits in the phase space of position and momentum $\{x,p\},$ once more as a harmonic oscillator but these orbits are now centered around the point $x=mc^2/\upsilon_{sc}.$ These closed and bound orbits exist both for the Dirac particles and antiparticles, implying that both are confined around the same trajectories.

\paragraph{Pseudoscalar potential and Dirac oscillator.-} The last analyzed case is the pseudo-scalar term $\upsilon_{ps}x\sigma_y$ with $\upsilon_{ps}=m\omega c.$ There is a similarity between
\begin{equation}
  H = c (\sigma_x p + \sigma_y m\omega x) + mc^2\sigma_z
\end{equation}
and the Hamiltonian of a $1+1$ Dirac oscillator~\cite{Bermudez07a},
\begin{equation}
  i\hbar\partial_t\psi = \left[
    c \vec{\alpha}(\vec{p} + i m\omega \beta \vec{x}) + mc^2\beta\right] \psi,
\end{equation}
but this coincidence only happens because of our restricted dimensionality, $\gamma^5=\alpha\beta.$ With the oscillator length scale, $a_{osc}=\sqrt{\hbar/m\omega}$, and the Rabi frequency  $\hbar\Omega=\sqrt{\hbar\omega mc^2},$ the pseudoscalar potential can be reformulated as a detuned Jaynes-Cummings model
\begin{equation}
  i\hbar\partial_t\psi = \left[\hbar\Omega(i \sigma_- a^\dag - i \sigma_+ a) + mc^2\sigma_z\right]\psi,
\end{equation}
which should be easy to simulate. A similar calculation as before produces again closed phase space orbits
\begin{equation}
  \frac{1}{m\omega}p^2 + x^2 m\omega + \hbar \sigma_z = E^2,
  \label{eq:orbits}
\end{equation}
which are a consequence of the JC discrete spectrum
\begin{equation}
  E_n = \pm mc^2\sqrt{ n \frac{\hbar\omega}{mc^2}+1} \sim mc^2+\frac{1}{2}n\omega\hbar.
\end{equation}
Note also that a replacement of $\omega\to-\omega$ is still a Dirac oscillator, but then it turns into an anti-JC Hamiltonian.

\paragraph{Conclusions and outlook.-}
We have proposed a quantum simulation of the $1+1$ Dirac equation with potentials in trapped ions, bringing together the physics of relativistic quantum mechanics to a controllable tabletop experiment in quantum optics. Our simulation protocols can be combined with tools for monitoring and preparing the quantum state of the ions. In particular, newly developed techniques ~\cite{Lougovski06,Gerritsma10a} make it possible to obtain the position, momentum and probability distribution $|\psi(x)|^2$ of the particle in an efficient way, allowing the ``frame-by-frame'' reconstruction of the scattering event. Besides this, the initial state of the Dirac particle can be accurately engineered in position, momentum and even energy branch~\cite{Gerritsma10a}. This is important for the electrostatic potential because, by preparing a particle with positive energy, we could see a full reflection for shallow slopes and full transmission for steeper slopes, making the Klein paradox visible.
As an example, similar to Ref.~\cite{Gerritsma10b}, one could prepare a positive energy wave packet with average momentum $\hat{p}=4\hbar/\Delta$ and choose experimentally accessible parameters $\tilde{\Omega}_b=\tilde{\Omega}_r=2\pi \times$~20~kHz, $\Omega_1=2\pi \times$~1~kHz, $\eta=0.05$ and $\tilde{\Omega}_2 = 2 \pi \times$~50~kHz. This permits to obtain a tunneling probability of 0.5 which could be observed in the lab. Numerical simulations show  that the whole tunneling dynamics with these parameters takes place within 1~ms, which is well within the motional and internal state coherence time. To study the Klein tunneling, see the case of electric potential and Fig.~\ref{fig:scattering}, the slope strength of the mass of the simulated particle can be varied by changing laser intensities and frequencies.

J.C. acknowledges funding from Basque Government BF108.211 and E.S. from UPV/EHU Grant GIU07/40, Spanish MICINN project FIS2009-12773-C02-01, Basque Government Grant IT472-10, EuroSQIP and SOLID European projects. J.J.G.-R. acknowledges funding from Spanish projects MICINN FIS2009-10061 and QUITEMAD. R.G. acknowledges support by the European Commission (Marie-Curie program).


\begin{thebibliography}{10}
\bibitem{Comment} The one-ion proposal for observing the Klein tunneling in Ref.~\cite{Lamata07} is flawed. A step potential in time, as proposed there, is not equivalent to the required potential in space.
\end{thebibliography}
\end{document}